\documentclass[sigconf]{acmart}

\usepackage[T1]{fontenc}
\usepackage{color}
\usepackage{xspace}
\usepackage{float}
\usepackage{listings}
\usepackage{amsmath}
\usepackage{relsize}
\usepackage{xspace}

\newcommand{\ourLib}{Haskell-TORCS\xspace}
\newcommand{\eg}{{\em e.g.~\xspace}}

\newcommand{\CC}{C\nolinebreak\hspace{-.05em}\raisebox{.4ex}{\tiny +}\nolinebreak\hspace{-.10em}\raisebox{.4ex}{\tiny +}}
\def\CC{{C\nolinebreak[4]\hspace{-.05em}\raisebox{.4ex}{\tiny ++}}\xspace}

\copyrightyear{2017}
\setcopyright{acmcopyright}
\acmConference[SCAV 2017]{2017 1st International Workshop on Safe Control of Connected and Autonomous Vehicles (SCAV 2017)}{April 2017}{Pittsburgh, PA USA}
\acmISBN{978-1-4503-4976-5/17/04}
\acmPrice{15.00}
\acmDOI{http://dx.doi.org/10.1145/3055378.3055385}

\begin{document}
\title{Vehicle Platooning Simulations with Functional Reactive Programming}

\author{Bernd Finkbeiner}
\orcid{1234-5678-9012}
\affiliation{%
  \institution{Saarland University}
  \state{Germany} 
}

\author{Felix Klein}
\orcid{1234-5678-9012}
\affiliation{%
  \institution{Saarland University}
  \state{Germany} 
}

\author{Ruzica Piskac}
\orcid{1234-5678-9012}
\affiliation{%
  \institution{Yale University}
  \state{CT, USA} 
}

\author{Mark Santolucito}
\orcid{0000-0001-8646-4364}
\affiliation{%
  \institution{Yale University}
  \state{CT, USA} 
}

\begin{abstract}
Functional languages have provided major benefits to the verification community.
Although features such as purity, a strong type system, and computational abstractions can help guide programmers away from costly errors, these can present challenges when used in a reactive system.
Functional Reactive Programming is a paradigm that allows users the benefits of functional languages and an easy interface to a reactive environment.
We present a tool for building autonomous vehicle controllers in FRP using Haskell.
\end{abstract}

%
%

\ccsdesc[500]{Computer systems organization~Embedded and cyber-\\physical systems}
\ccsdesc[300]{Software and its engineering~Embedded software}
\ccsdesc[300]{Software and its engineering~Real-time systems software}

\keywords{FRP, Autonomous Vehicles}

\setcounter{footnote}{1}
\newcounter{x}
\def\mynewline{\ifnum\value{footnote} > \value{x} \\ \fi}
\title{Vehicle Platooning Simulations with Functional Reactive Programming}
\maketitle

\title{Vehicle Platooning Simulations with Functional Reactive Programming}

\section{Introduction}

Autonomous vehicles are considered to be one of the most challenging
types of reactive systems currently under development~\cite{AlurMT16,
  WongpiromsarnKF11, RamanDSMS15}. They need to interact reliably with
a highly reactive environment and crashes cannot be tolerated.  Life
critical decisions have to be made instantaneously and need to be
executed at the right point in time.

The development of autonomous vehicles and other cyberphysical systems
is supported by a wide spectrum of
programming and modeling methodologies, 
including synchronous programming languages like Lustre~\cite{conf/popl/CaspiPHP87} 
and Esterelle~\cite{conf/concur/BerryC84},
hardware-oriented versions of imperative programming languages like
SystemC~\cite{open2006ieee}, and visual languages like MSCs and Stateflow-charts~\cite{harel2003message,journals/scp/Harel87}.
The question of which programming paradigm is best-suited to write
easy-to-understand, bug-free code is still largely unresolved.

In the development of other forms of critical software, outside the
embedded domain, developers increasingly turn to functional
programming (cf.~\cite{frankau2009commercial}).  The strong type system in functional
languages largely eliminates runtime errors~\cite{cardelli1996type}.
Higher-order functions like \texttt{map} often eliminate the need for
explicit index counters, and, hence, the risk of ``index out of
bounds'' errors.  Functional purity reduces the possibility of
malformed state that can cause unexpected behavior.

While mathematical models of embedded and cyberphysical systems often
rely on functional notions such as stream-processing
functions~\cite{series/mcs/BroyS01,conf/csdm/Broy12}, the
application of functional programming in the practical development of
such systems has, so far, been limited. One of the most advanced
programming language in this direction is Ivory, which was used in the
development of autonomous vehicles~\cite{pike2014}.  Ivory is a
restricted version of the C programming language, embedded in Haskell.
It provides access to the low level operations necessary for embedded
system programming, but still enforces good programming practice, such
as disallowing pointer arithmetic, with a rich type system.

Ivory does not, however, have an explicit notion of time.
It cannot deal directly with the integration of 
continuous and discrete time, which is fundamental for the
development of a cyberphysical system. For example,
in a car, continuous signals, such as the velocity or acceleration,
mix with the discrete steps of the digital controller.

In this paper, we investigate the use of functional programming
in a domain where the interaction between continuous and discrete signals
is of fundamental importance. We build a vehicle controller capable
of both autonomous vehicle control and multi-vehicle communication,
such as the coordination in platooning situations.

Our approach is based on Functional Reactive Programming
(FRP)~\cite{hudak2003arrows,hudak2000haskell}.
The fundamental idea of FRP is to extend the classic building blocks 
of functional programming (\eg monads, arrows,
applicatives)
with the abstraction of a \emph{signal} to
describe time-varying values. FRP programs can be exceptionally
efficient.  For example, a network controller recently implemented as
an FRP program on a multicore processor outperforms any other such
controller existing today~\cite{Voellmy:2012:SSD:2377677.2377735}.

We have built a library, \ourLib, to use FRP to control a vehicle inside a simulation.
The library interfaces Haskell FRP programs to TORCS, The Open Racing Car Simulator, an open-source vehicle simulator~\cite{torcs}.
TORCS has been used in the Simulated Car Racing Championship competition~\cite{SCRC}, as well as other autonomous vehicle research projects~\cite{xu2016experimental,OnievaPAMP09,conf/cig/CardamoneLL09,conf/cig/MunozGS10}. 
Through \ourLib, the Haskell program has access to the sensors and actuators of the car, as well
as communication channels between different vehicles.
Such a simulator is a  critical component of modern autonomous vehicle research, especially towards the goal of safe platooning algorithms~\cite{kamali2016formal}.

\begin{figure}[t]
\includegraphics[width=0.45\textwidth]{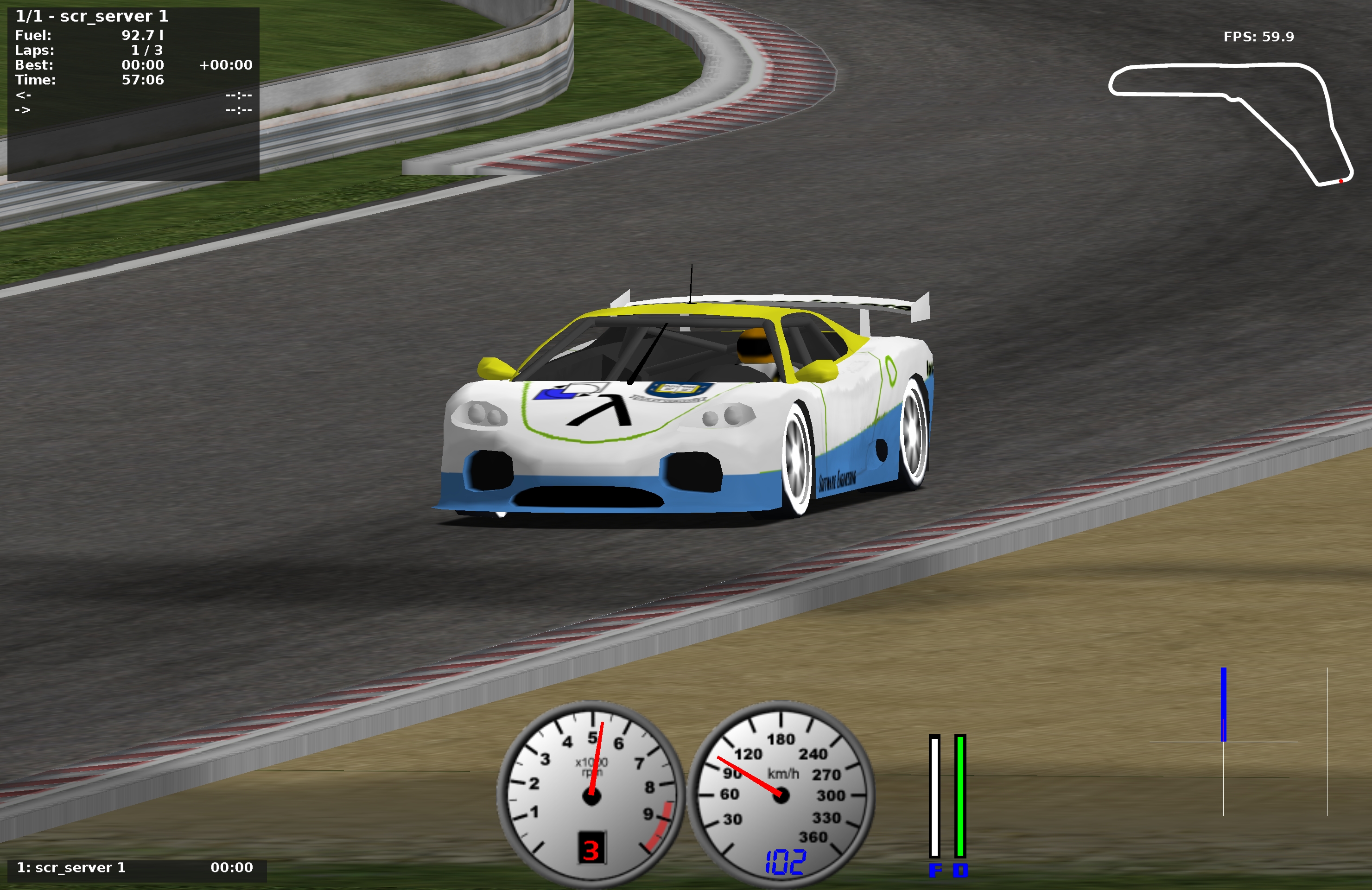}
\caption{A screenshot of Haskell controlling the autonomous vehicle in the TORCS simulator.}
\label{fig:race}
\end{figure}

We report on our experience with two case studies, one in which we implement a controller for a solo car, and another for multi-vehicle platooning using a communication channel between the cars.
Our controller successfully navigates the TORCS preloaded tracks with reasonable speed and finesse while avoiding collisions (see Fig.~\ref{fig:race}).
Furthermore, the case study illustrates that the functional approach indeed leads to elegant, easily understandable, and safe code.
The ability to run full simulations for solo and platooning vehicles is a critical piece to advancing the state of the art in using FRP for autonomous vehicle control.

\section{FRP}

The most common solution for the construction of reactive systems in an imperative setting are call-back frameworks, embedded into a loop.
The call-backs are either used to query the state of variables, or to change them.
This imperative approach is well suited for rapid prototyping of small systems.
However, tracing behaviors over time quickly becomes unmanageably complex for larger systems.

Functional Reactive Programming instead introduces a concrete abstraction of time that allows the programmer to safely manipulate time-varying values. 
The key abstraction is given by a \textit{signal}, providing the programmer with a simple type interface:

\begin{lstlisting}
  type Signal a = Time -> a
\end{lstlisting}

For example, the type \texttt{Signal Image} represents a video, while \texttt{Signal Steer} captures a steering wheel operated over time.
To better understand how our library works, we now introduce the basic concepts and terminology from FRP.

\subsection{Arrowized FRP}

There are many types of FRP based on different abstractions from type theory.
Expressive abstractions, such as monads, allow for complex manipulation of signal flows~\cite{van2014monadic}. 
However, for most applications they are far too expressive.
We instead focus on an FRP library, Yampa, which uses the arrow abstraction, or so called Arrowized FRP~\cite{hudak2003arrows}.
Arrows generally run faster and with little need for manual optimization~\cite{yallop2016causal}, but are fundamentally less expressive than a monadic FRP~\cite{lindley2011idioms}.
This more restrictive language is in fact a benefit, as it makes it harder for the programmer to introduce errors.
As we will see in the sequel, Yampa is still powerful enough to write complex controllers to drive an autonomous vehicle, or even to communicate with other vehicles.
At the same time, the syntax is clear and accessible enough to make for an easy introduction to the FRP paradigm.

Along with signals, Yampa also introduces the abstraction of a \textit{signal function (SF)}.
This is a transformer from one signal to another.

\begin{lstlisting}
  type SF a b = Signal a -> Signal b
\end{lstlisting}

\noindent Using the previous signals, imagine a type for a steering function, which operates based on a video stream, such as

\begin{lstlisting}
  turn :: SF Image Steer
\end{lstlisting}

\noindent This function processes video and uses it to decide how to steer.
We omit an implementation, as the details of the data transformation are not relevant to the structure of the FRP code.

Haskell provides special syntax for Arrowized FRP, which mimics the structure of control flow charts.
The syntax provides a composition environment, in which the programmer just manages the composition of arrow functions.
Inputs are read in from the right hand side, and piped to the left hand side (\texttt{output <- function -< input}).
A demonstration is given in Listing~\ref{lst:arrows}.

The example introduces 
\begin{lstlisting}
  avoid :: (Image, Steer) -> Steer
\end{lstlisting}
a pure function that adjusts the basic steering plan based on the image to avoid any obstacles.
In Listing~\ref{lst:arrows}, this \texttt{avoid} function is lifted to the signal level using:
\begin{lstlisting}
  arr :: (a -> b) -> SF a b
\end{lstlisting}
The function \texttt{turn} is already on the signal level (has an SF type). Hence, we do not need to lift it.

\begin{lstlisting}[float,floatplacement=h!,caption=Basic Arrowized FRP syntax,label=lst:arrows]
myDriver :: SF Image Steer
myDriver = proc image -> do
  basicSteer    <-     turn  -< image
  adjustedSteer <- arr avoid -< (image, basicSteer)
  returnA -< adjustedSteer
\end{lstlisting}

\subsection{Stateful FRP}

To avoid obstacles on the road, we might write an \texttt{avoid2} function as shown in Listing~\ref{lst:loop}, which requires two images to calculate the adjusted steering command. 
For this, we need a mechanism to maintain state between each processing step.
Two images would be necessary to filter noise in the image, or calculates the velocity of an approaching obstacle.
To implement it, we use an abstraction called \texttt{ArrowLoop} to  save the previous state of the image for the next processing step.
The syntax is presented in Listing~\ref{lst:loop}.
Intuitively, \texttt{ArrowLoop} gives us a recursive computation, as also indicated by the \texttt{rec} keyword\footnote{We elide the technical details for the purposes of this presentation and refer the interested reader to~\cite{paterson2001icfp}.}.

\begin{lstlisting}[float,caption=Using ArrowLoop to send feedback,label=lst:loop]
myDriver :: SF Image Steer
myDriver = proc image -> do
  rec
    oldI          <- iPre null  -< image
    basicSteer    <-      turn  -< image
    adjustedSteer <- arr avoid2 -< (image, oldI, basicSteer)
  returnA -< adjustedSteer
\end{lstlisting}

The predefined function \texttt{iPre} takes an initial state, in our case an empty image, and saves images for one time step, each time it is processed.
This way, we create a feedback loop that is then used in the updated \texttt{avoid} function.
At the same time, the \texttt{rec} keyword is used to denote a section of arrow code with  mutual dependencies\footnote{Without the keyword, there is an unresolvable dependency loop.}.

\section{\ourLib}
TORCS, The Open Racing Car Simulator, is an existing open source vehicle simulator~\cite{torcs} that has bindings for various languages~\cite{SCRC}.
We provide the first bindings for Haskell, and further extend this into a full library for multi-vehicle simulations.
The library is an open source library, called \ourLib, and publicly available at \url{https://hackage.haskell.org/package/TORCS}.
We now explain the functionality provided by our library, and highlight the ability of FRP to create modular and flexible controllers with clean code for autonomous vehicles.

\subsection{Basics}

To interface with \ourLib, a user must implement a controller that will process the \texttt{CarState}, which contains all the data available from the sensors.
The controller should then output a \texttt{DriveState}, which contains all the data for controlling the vehicle.
This transformation is succinctly described as the now familiar \textit{signal function}.
The core functionality of \ourLib is captured in the function \texttt{startDriver}, which launches a controller in the simulator.
This function automatically connects a \texttt{Driver} to TORCS, which results in continuous \texttt{IO()} actions, the output type of this function.

\vspace{0.2em}
\begin{lstlisting}
  type Driver = SF CarState DriveState
  startDriver :: Driver -> IO ()
\end{lstlisting}
\vspace{0.2em}

\noindent The sensor and output data structures contain all the typical data available in an autonomously controlled vehicle.
\texttt{CarState} includes fields like \texttt{rpm} to monitor the engine, or \texttt{track} to simulate an array of LiDAR sensors oriented to the front of the vehicle.
\texttt{DriveState} includes fields like \texttt{accel} to control the gas pedal, or \texttt{steering} to control the angle of the steering wheel.
A full description of the interface is available in the Simulated Car Racing Competition Manual~\cite{SCRCManual}. 

\subsection{Case Study : Driving}

As a demonstration of the \ourLib library in use, we implemented a simple controller, shown in Listing~\ref{lst:driver}. 
The code is complete and immediately executable as-is together with an installation of TORCS.
Our controller successfully navigates, with some speed and finesse, a vehicle on track, as shown in Fig.~\ref{fig:race} along with a video demonstration\footnote{\url{http://www.marksantolucito.com/torcsdemo}}.
The controller uses \texttt{ArrowLoop} to keep track of the current gear of the car.
Although the gear is available as sensor data, it is illustrative to keep track locally of this state.
In general, the \texttt{ArrowLoop} can be used to maintain any state that may be of interest in a future processing step.
Additionally, notice all of the data manipulation functions are pure, and lifted via the predefined function~\texttt{arr}.

One major advantage of FRP is this separation of dependency flow and data level manipulation. 
This abstraction makes it possible to easily reason about each of the components without worrying about confounding factors from the other.
For example, if a programmer wants to verify that the steering control is correct, it is semantically guaranteed that the only function that must be checked is \texttt{steering}.
Because of Haskell's purity, this is the only place where the steering value is changed. This significantly reduces the complexity of verification or bug tracking in case of an error.

\begin{lstlisting}[float,floatplacement=TR,caption=A complete basic controller in Yampa, label=lst:driver,framesep=0pt,rulesep=0pt,frame=lines,framerule=0pt]
{-# LANGUAGE Arrows, MultiWayIf, RecordWildCards #-}
module TORCS.Example where
import TORCS.Connect
import TORCS.Types

main = startDriver myDriver

myDriver :: Driver
myDriver = proc CarState{..}  -> do
  rec 
    oldG <- iPre 0 -< g
    g <- arr shifting -< (rpm, oldG)
    s <- arr steering -< (angle, trackPos)
    a <- arr gas -< (speedX, s)
  returnA -< defaultDriveState 
    { accel = a, gear = g, steer = s }

shifting :: (Double, Int) -> Int
shifting (rpm, g) = if 
  | rpm > 6000 -> min 6 (g + 1)
  | rpm < 3000 -> max 1 (g - 1)
  | otherwise  -> g
 
steering :: (Double, Double) -> Double
steering (spd, trackPos) = let
  turns = spd * 14 / pi
  centering = turns - (trackPos * 0.1)
  clip x = max (-1) (min x 1)
 in
  clip centering

gas :: (Double, Double) -> Double
gas (speed, steer) = 
  if speed < (100 - (steer * 50)) then 1 else 0
\end{lstlisting}
%
\subsection{Case Study : Communication for Platoons}

Thanks to functional languages' exceptional support for parallelism, controlling multiple vehicles in a multi-threaded environment is exceedingly simple. 
In our library API, the user simply uses \texttt{startDrivers} rather than \texttt{startDriver}, and passes a list of \texttt{Driver} signal functions ``driving'' together.
In this way, we easily let various implementations race against each other, or build a vehicle platooning controller.
In the latter, the user can even extend the implementation to simulate communication between the vehicles.

Our library already provides a simple interface for simulating communication between vehicles.
In order to broadcast a message to the other vehicles in the simulation, the controller simply writes a message to the \texttt{broadcast} field of \texttt{DriveState}.
That message is then sent to all other vehicles as soon as possible, and received in the \texttt{communication} field of the input \texttt{CarState}.

A fragment of communication code is given in Listing~\ref{lst:platoon}, to pass messages between vehicles. 
In this fragment, a vehicle checks if a collision is imminent, and can request for the other cars in the platoon to go faster and move out of the way.
Every vehicle also checks if any other car has requested for the platoon to speed up, and will adjust its own speed accordingly.
These functions can be added to a controller, like the one in Listing~\ref{lst:driver}, with little effort.

We allow all vehicles in the simulation to communicate irrespective of distance and with zero packet loss.
However, users are free to implement and simulate unreliable communications, or distance constraints.  

\subsection{Implementation} 

\mbox{\ourLib} uses Yampa~\cite{courtney2003yampa} as the core FRP library, though its structure can easily be adapted to any other Haskell FRP library.

TORCS uses a specialized physics engine for vehicle simulations, that includes levels of detail as fine grained as tire temperatures effect on traction. 
When TORCS is used in the Simulated Car Racing Championship competition~\cite{SCRC}, each car is controlled via a socket that sends the sensor data from the vehicle and receives and processes the driving commands.
So too, \ourLib communicates over these sockets to control vehicles inside the TORCS simulations.

In addition to the core controller functionality, we have also augmented \ourLib with the ability to test vehicle platooning algorithms that utilize cross-vehicle communication.
The communication channels are realized via a hash map, using the \texttt{Data.Hashmap} interface, from vehicle identifiers to messages.
Each vehicle is given write permissions to their unique channel, where all other vehicles have read-only permissions.
The access is mutually exclusive, which is ensured by Haskell's \texttt{MVar} implementation, a threadsafe shared memory library.
This ensures that there will never be packet loss in the communication.
 
\begin{lstlisting}[float,floatplacement=H,caption=Communicating between controllers, label=lst:platoon,frame=lines,framerule=0pt,framesep=0pt,rulesep=0pt]
request :: Double -> Message
request dist = 
  if dist < 3 then "faster" else ""

adjustSpeed :: (Communications, Double) -> Double
adjustSpeed (comms, oldSpeed) =
  if any (map (== "faster") comms) then s + 10 else s
\end{lstlisting}

\section{Related Work}

TORCS has been proven to provide an expressive framework for the research community~\cite{OnievaPAMP09,conf/cig/CardamoneLL09,conf/cig/MunozGS10}. 
Notably, it has even been used for formal verification of platoons~\cite{kamali2016formal,xu2016experimental}. 
None of these works have used FRP as the language for the controller.
With the assistance of FRP, we build vehicle controllers in a principled way that allows users to manipulate sensor data in a transparent and well structured environment.

To the best of our knowledge this is the first FRP-based vehicle simulator.
Although there are many bindings to various vehicle simulators, these tend to use imperative languages.
For instance, TORCS allows users to directly edit the source code and add a new car in \CC.
There are also TORCS bindings for Python, Java, and Matlab, which have been used in the SCRC competition~\cite{SCRC}.

FRP specifically has been proposed as a tool for vehicle control~\cite{kazemi2016,zou2016}, where FRP was extended to prioritize functions for timing constraints. However, due to the lack of a compatible simulator, the vehicle simulation never was implemented. 
FRP has also been used for embedded systems~\cite{helbling2016juniper} and networking~\cite{voellmy2012scalable}.
The FRP networking library took advantage of Haskell's multicore support and significantly outperformed competing tools written in \CC and Java.

The videogame Grand Theft Auto (GTA)~\cite{gtaV} has also been used to train image recognition software for autonomous vehicles~\cite{gtaPrinceton}.
While GTA is a professionally produced game with more attractive graphics, it is proprietary software not designed for autonomous vehicle research.
The only available sensor data are gameplay images, which are a limited model for autonomous vehicles.
Using GTA as a meaningful control simulator would still be a valuable tool, but we leave this to future work.

\section{Conclusions}

We have presented a library to write autonomous vehicle controllers in FRP that supports cross-vehicle communication.
This work opens the door for further research in using the powerful FRP paradigm for building safer, more reliable controllers for one of the most critical applications in reactive systems.


\subsection*{Acknowledgments} 
Supported by the European Research Council (ERC) Grant OSARES (No.\ 683300) and 
 by the National Science Foundation (NSF) Grant CCF-1302327.

\bibliographystyle{ACM-Reference-Format}
\bibliography{sigproc} 

\end{document}